# Wavelength- and time-division multiplexing via pump current variation of a pulsed semiconductor laser – a method of synchronization for quantum key distribution

Roman Shakhovoy, Marius Puplauskis, Violetta Sharoglazova, Elizaveta Maksimova, Selbi Hydyrova, Vladimir Kurochkin, and Alexander Duplinskiy

*Abstract*— The dependence of the semiconductor laser wavelength on the pump current is a well-known phenomenon, which is generally attributed to a change in the refractive index of the active layer due to carrier injection. This effect is usually considered to be a drawback as it causes frequency chirping of the pulses produced via direct current modulation. Here, we show that at high values of thermal resistance of a laser diode, the lasing wavelength is red-shifted due to the change of the refractive index caused by a thermal effect and this shift may significantly exceed (in absolute value) the blue shift related to a refractive index change induced by the carrier injection. We propose to benefit from the ability to tune wavelength by the pump current variation and use the same laser to generate qubits and synchronization optical pulses for quantum key distribution at different wavelengths. To demonstrate the proposed method of synchronization, we perform numerical simulations as well as proof-of-principle experiments.

*Index Terms*— Semiconductor lasers, temperature effects, wavelength-division multiplexing, synchronization, quantum key distribution.

## I. INTRODUCTION

QUANTUM key distribution (QKD) is a perspective technology towards unconditionally secure communication systems. QKD allows two considerably distant parties – the sender (Alice) and the receiver (Bob) – share cryptographic keys through an untrusted channel. The principal feature of QKD is that the secrecy of shared keys is based on fundamental laws of quantum physics [1]. Despite the successful implementation of QKD systems around the world, they still should overcome several technical difficulties to achieve commercial success. The main challenges include a raise of the key transmission rate, an increase of the distance between communicating parties [2-5], an effective temperature stabilization of optical schemes (at least in some implementations [1]), as well as an effective way of the sender and receiver synchronization [2, 3, 6-8]. The latter issue is often not discussed in detail, especially when it comes to the implementation of the QKD system in laboratory conditions; however, the synchronization problem is extremely important in real systems.

Generally, the sender and receiver devices have their own reference frequency generators, which operate independently and thus accumulate a relative error over time. Therefore, regular measurements of the phase difference between the generators are required to ensure maintenance of synchronous system operation. The measurements may be performed by transferring from time to time the calibration signal from one of the devices to another for frequency comparison. In early laboratory implementations, synchronization was achieved with electrical signals, e.g., via a coaxial cable connecting the devices. In more realistic applications, Alice and Bob exchange optical signals for long-distance clock synchronization [2, 9]. Obviously, it is difficult to use weak optical signals to synchronize the communication system, since the probability of their arrival at Bob's detector is small (especially at large lengths of the quantum channel) and the time of their appearance is irregular. Therefore, synchronization is usually performed with strong optical pulses transmitted through a synchronization channel, which may be implemented in a separate fiber [10], in a free-space optical link [9], or may share the same fiber with quantum states [11-14] by means of wavelength-division multiplexing (WDM) or time-division multiplexing (TDM). The latter option is generally preferred as it is not always possible to provide a separate fiber for synchronization purposes due to cost issues or lack of available fiber capacity [15].

Some implementations of QKD provide an opportunity to use a *single* laser to generate both quantum and synchronization signals. Thus, in the plug-and-play scheme [1, 16], Bob sends to Alice strong optical pulses, which are used directly for synchronization. These pulses are then attenuated and converted into weak quantum signals, which Alice sends back to Bob. A single laser can be also used in continuous-variable (CV) QKD [17], where quantum and synchronization signals are measured with the same (classical) detector. In [18], authors considered the CV-QKD scheme, where a semiconductor laser randomly emitted both types of signals mixed in time. These mixed pulses were then sent together to Bob, where one type of pulses was separated from the other on the fly, thus realizing synchronization.

Most implementations of common QKD protocols (BB84 [19-21], MDI-QKD [22], COW [23]) require, however, *separate* laser sources to generate quantum and synchronization pulses, since different detectors are required to detect these two types of pulses, i.e., the signals must be physically separated in space. Such separation can be carried out with WDM; however, simultaneous propagation of strong and weak pulses over a common fiber is accompanied by the appearance of secondary photons, resulting from Raman and other nonlinear interactions of the intense synchronization signal with the fiber [19]. These secondary photons can lead to significant spurious illumination of single photon detectors and, consequently, to an increase of the quantum bit error rate (QBER). Therefore, to avoid crosstalk between synchronization and quantum signals, TDM is generally used in addition to WDM [13, 19].

There are other solutions that allow avoid the use of an additional laser subsystem required for synchronization. They include algorithms that employ public information about quantum states to correct clock drift and time offset [6-8, 24, 25]. Such procedures are called *qubit-based protocols* since they use the same qubits exchanged during the QKD protocol to implement synchronization and do not generally require additional hardware. These methods make use of correlations between qubits emitted by Alice and received by Bob. In [7], a fast synchronization algorithm based on measuring the arrival time of qubits was developed. It was tested experimentally in the three-polarization-state version of the BB84 system with a 26 km long fiber quantum channel [24] and the existing telecommunication infrastructure [25] and showed high stability and record-low QBER. In [6], the authors introduced a Bayesian probabilistic algorithm that engaged all published information (the basis and decoy state choices, mean photon number choices, Bob's detection events) to find the clock offset without sacrificing key generation rate efficiently. The performance of this algorithm was demonstrated through numerical simulations with a model of the three-state BB84 protocol with decoy states. Similar qubit-based synchronization protocols for entanglement-based QKD systems were also introduced [8, 26-30]. Thus, the method from [8] was based on measuring the correlation of Alice's and Bob's detection events, such that the key generation rate was not diminished. However, these qubit-based methods do not allow simple adjustments. Moreover, these procedures are not trivial to implement and have limitations in their application.

In this paper, we propose a time and wavelength division multiplexing (TWDM) method of synchronization with a single laser source, which can be employed in QKD systems, where an additional laser is generally required for this purpose. In our method, TDM is implemented by dividing signals into alternating trains of synchronization and information pulses, whereas WDM is based on the lasing frequency shift due to variation of the pump current. The main advantage of the proposed method is reduction of required hardware in Alice's device and cost reduction of the QKD system. To our knowledge, this approach has not been previously discussed in the context of QKD synchronization.

We also demonstrate here that the laser frequency shift induced by a change of the pump current is governed mainly by the heating of the active layer and not by the chirp effect. We consider in detail the mechanism of the temperature change inside the active layer and develop a simple experimental method for evaluating the effective thermal resistance of a laser diode, a parameter responsible for heating efficiency.

## II. GENERAL DESCRIPTION OF A SYNCHRONIZATION SYSTEM

As discussed in the introduction, the specific implementation of the synchronization protocol depends on a particular QKD scheme as well as the specific hardware used in the system. However, we will generalize here the synchronization protocol without going into details. We will assume that the QKD system employs the optical scheme sketched in Fig. 1(a). Although the scheme in the figure is assumed to be implemented in a fiber, one can always substitute some fiber-optic components with their bulk counterparts and use this scheme, e.g., for free space communication.

Alice in Fig. 1(a) contains a laser (L1), which generates identical laser pulses grouped into trains. The trains pass through the encoder, which prepares weak coherent pulses (quasi-single-photon states) according to the selected QKD protocol. Prepared quantum states are then sent to Bob through an optical isolator, which prevents laser seeding attack [31].

Bob contains the corresponding decoder followed by single photon detectors. To correctly interpret registered clicks on Bob's side, it is necessary to prevent a mismatch between clocks of Alice and Bob. Incorrect matching between laser pulses sent by Alice and the modulation signal driving Bob's decoder will increase the number of errors or won't let them generate the key at all. Therefore, one of the clock generators (generally, Bob's one) must have a mechanism for fast frequency tuning (the tuning range should be sufficient to compensate for any possible relative detuning).

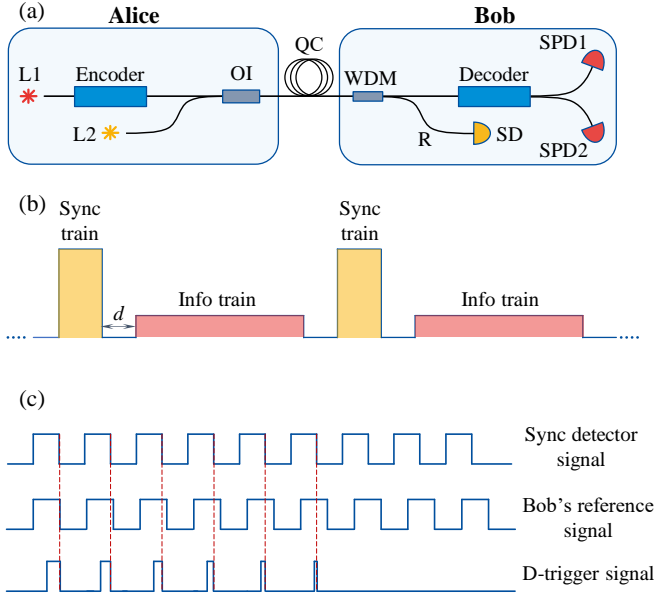

Fig 1. (a) Simplified optical scheme for quantum key distribution: L1 – principal laser, L2 – sync laser, OI – optical isolator, SPD1, SPD2 – single photon detectors, SD – sync detector, QC – quantum channel, R – reflection arm of the WDM filter. (b) Schematic plot of laser pulse trains. (c) Schematic representation of a (digitized) signal of sync detector and Bob's reference signal with their comparison at D-trigger.

For synchronization, Alice in Fig. 1(a) uses a sync laser (L2), whose central frequency is detuned from L1. Sync laser inserts a train of pulses in front of each L1 train as shown in Fig. 1(b); these pulses are not attenuated (the envelope of the sync train has therefore a higher amplitude in the figure) and can be thus detected by Bob with a conventional photodetector, which we call the sync detector (SD). To separate trains with qubits from synchronization trains, one can use WDM, e.g., a conventional WDM-filter equipped with the reflection arm (such a configuration is assumed in Fig. 1(a)) or the two optical bandpass filters (one for the qubits and the other for synchronization pulses) accompanied by additional couplers or beam splitters (see Fig. 2). SD converts synchronization pulses into an electrical signal, which after digitization is compared with the reference signal generated by Bob's frequency generator (Fig. 1(c)). The comparison can be performed, e.g., with the D-trigger. The signal from the D-trigger is shown on the bottom line of Fig. 1(c) (a significant mismatch between compared frequencies is assumed for clarity). One can see that when the phase shift between signals exceeds 90 degrees, a continuous stream of 'zeros' comes from the trigger. Calculating the number of 'ones', Bob may thus determine the detuning. Several thousand pulses in the synchronization train are generally enough to measure the mismatch between the compared signals with sufficient accuracy. Bob then adjusts the frequency of his clock, eliminating the difference with Alice's clock.

The method we propose allows to remove the laser L2 from Alice's optical scheme and use the same laser to generate both synchronization and qubit trains. It is done by repeatedly switching the laser L1 from information to synchronization pulse generation mode. Possible implementations of Alice's optical scheme with such type of synchronization are sketched in Fig. 2. In Fig. 2(a), Alice is equipped with two individual filters for synchronization and information signals, whereas in Fig. 2(b), she uses the WDM-filter equipped with the reflection arm. (In principle, it is irrelevant which of the arms – reflection or transmission – is used for the sync signal; therefore, we use the notation R/T in Fig. 2(b).) Although these two implementations are functionally equivalent, they have a significant difference in terms of system security. This problem will be discussed below.

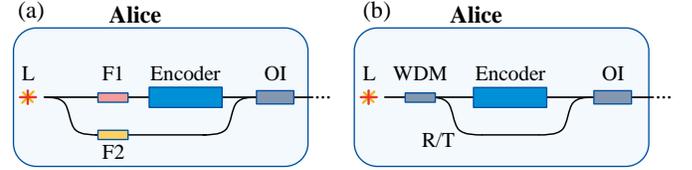

Fig. 2. Possible implementations of Alice's optical scheme with a single laser generating both info and sync pulse trains. F1, F2 – optical bandpass filters. Other notations are the same as in Fig. 1.

Switching the laser from the synchronization mode to generation of information pulses can be performed by changing the bias current on the laser. In fact, sufficient change in the pump current shifts the lasing frequency far enough for the trains to be separated via spectral filtering. Thus, the synchronization protocol remains essentially unchanged, whereas Alice becomes simpler and cheaper. Yet two questions arise here: 1) what is the frequency shift mechanism and 2) how quickly the lasing frequency is established after switching the current. These questions are of the main scientific interest and the answer to them is the leitmotif of this work.

## III. RATE EQUATION ANALYSIS

To study the laser frequency shift caused by the change in the pump current, we start from the system of semiconductor laser rate equations [32, 33]:

$$\dot{N} = I/e - N/\tau_e - QG/(\Gamma\tau_{ph}),$$
$$\dot{Q} = (G-1)Q/\tau_{ph} + C_{sp}N/\tau_e, \quad (1)$$
$$\dot{\varphi} = \alpha(G_L-1)/(2\tau_{ph}),$$

where $Q$ is the normalized electric field intensity corresponding to the photon number inside the laser cavity and related to the output power by $P = Q\eta\hbar\omega_0/(2\Gamma\tau_{ph})$, where $\hbar\omega_0$ is the photon energy ($\omega_0$ is the central angular frequency), $\eta$ is the differential quantum output, $\Gamma$ is the confinement factor, $\tau_{ph}$ is the photon lifetime inside the cavity, and the factor 1/2 takes into account that the output power is measured only from one facet. Onwards, $\varphi$ is the phase of the field, $N$ is the carrier number, $I$ is the pump current, $e$ is the absolute value of the electron charge, $\tau_e$ is the effective lifetime of the electron, the factor $C_{sp}$ corresponds to the fraction of spontaneously emitted photons that end up in the active mode, $\alpha$ is the linewidth enhancement factor (the Henry factor [34]), and the dimensionless linear gain $G_L$ is defined by $G_L = (N - N_{tr})/(N_{th} - N_{tr})$, where $N_{tr}$ and $N_{th}$ are the carrier numbers at transparency and threshold, respectively. The gain saturation [35] is included in (1) by using the relation $G = G_L/\sqrt{1 + 2\chi P}$ or, equivalently, by $G = G_L/\sqrt{1 + 2\chi_Q Q}$, where $\chi$ is the power-related gain compression factor (in W$^{-1}$) [32, 35, 36] and $\chi_Q = \chi \eta \hbar \omega_0 / (2\Gamma \tau_{ph})$ is its dimensionless counterpart.

Note that conventional rate-equation description (1) is not entirely valid in the analysis of a DFB laser since it does not take into account the nonuniform distribution of the optical field along the cavity that appears in the longitudinal spatial hole burning (LSHB) effect. A more accurate description is based on the analysis of dynamic coupled-wave equations [37, 38]. However, the dependence of the lasing frequency on temperature considered in this article should be described in approximately the same way in both approaches. Moreover, this dependence is not related to the LSHB effect, so the choice of the model is not very significant. In addition, our goal is to clarify the physical nature of the lasing frequency shift caused by the pump current variation rather than to give the most accurate quantitative description of this phenomenon. For this purpose, conventional laser rate equations are much better suited than dynamic coupled-wave equations.

A. *Steady-state frequency shift*

A change in the pump current causes a change in the carrier concentration in the active layer, which leads to a change of its refractive index and, as a result, to a change of the lasing frequency. The corresponding frequency shift can be estimated with the formula [33, 39]:

$$\Delta\omega = \frac{\alpha}{2}\left[\frac{1}{P(t)}\frac{dP(t)}{dt} + \frac{\chi}{\tau_{ph}}P(t)\right], \quad (2)$$

which can obtained from (1) by neglecting the spontaneous emission term in the equation for $Q$ and substituting $G_L - 1$ from $\dot{Q}$ to $\dot{\varphi}$. At continuous wave (CW) operation, the derivative $dP/dt$ is zero, so, the frequency shift (2) becomes just proportional to the output power $P$: $\Delta\omega = \alpha\chi P/(2\tau_{ph})$. Output optical power can be well approximated by the following formula:

$$P = \frac{\eta \hbar \omega_0}{2e}(I - I_{th}), \quad (3)$$

where $I_{th}$ is the threshold current. Thus, we obtain the following relation for the frequency shift $\Delta\omega$:

$$\Delta\omega \approx \frac{\alpha \chi \eta \hbar \omega_0}{4e\tau_{ph}}(I - I_{th}). \quad (4)$$

The change of the pump current from some value $I_1$ to $I_2$ will thus cause the frequency shift

$$\Delta\omega_c = \Delta\omega_2 - \Delta\omega_1 = \frac{\alpha \chi \eta \hbar \omega_0}{4e\tau_{ph}}\Delta I, \quad (5)$$

where $\Delta I = I_2 - I_1$. (For a more discussion on the validity of (4) and (5) and benefits of these useful relations in the context of optical injection see [40].) Using typical values of semiconductor laser parameters from the following ranges: $\alpha = 2...5$, $\eta = 0.2...0.5$, $\chi = 5...50$ W$^{-1}$, $\tau_{ph} = 1...2$ ps, the central frequency $\omega_0/2\pi = 193.548$ THz ($\lambda_0 = 1550$ nm) and $\Delta I = 10$ mA, we will find: $\Delta\omega_c \approx 0.3...40$ GHz. The corresponding shift in the wavelength is $\Delta\lambda \approx -\lambda_0^2 \Delta\omega_c/(2\pi c) = -0.0024...-0.32$ nm. (Note that according to (5), $\Delta\omega_c > 0$ when $\Delta I > 0$.)

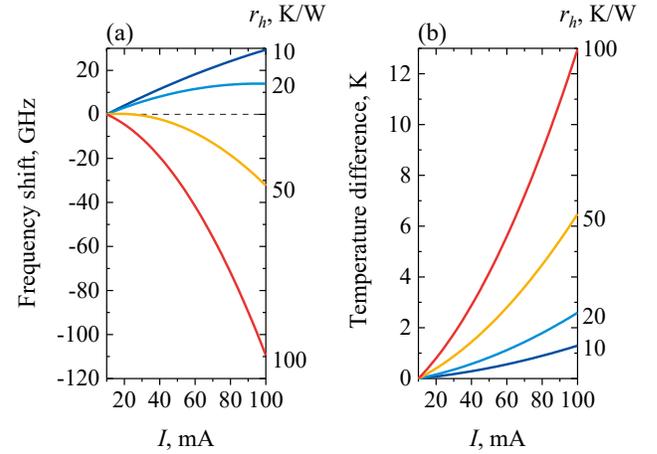

Fig. 3. (a) Theoretical dependences of the total laser frequency shift $\Delta\omega_{tot}$ on the pump current $I$ at different values of the thermal resistance $r_h$. (b) Corresponding dependences of the temperature difference $\Delta T$. In simulations, it was assumed that $I_{th} = 10$ mA, $R_L = 8$ Ohm, $V_g = 0.8$ V, $\eta = 0.3$, $\chi = 20.8$ W$^{-1}$ ($\chi_Q = 10^{-6}$), $\Gamma = 0.2$, $\tau_{ph} = 2$ ps, $\kappa_\omega/2\pi = 12$ GHz.

The frequency shift (5) is intrinsic for the semiconductor laser rate equations (1); therefore, there is no need to make additional efforts to calculate it. However, a change in the refractive index induced by the carrier injection is not the only reason for the lasing frequency shift. When an electric current passes through a laser diode, a resultant increase in temperature within the laser volume is observed, among other effects. This temperature rise leads to an additional spectral shift $\Delta\omega_T$. It should be included into laser rate equations by modifying the equation for the phase as follows:

$$\dot{\varphi} = \alpha(G_L - 1)/(2\tau_{ph}) - \kappa_\omega \Delta T, \quad (6)$$

where the temperature difference $\Delta T$ is the difference $T_2 - T_1$ between the steady-state temperatures at two different pump currents $I_2$ and $I_1$. The temperature coefficient $\kappa_\omega$ is related to several factors including the changes 1) in the width of the energy gap of the semiconductor material, 2) in the value of the refractive index ($\partial n/\partial T$), and 3) in the cavity length $L$ (or in the size of the Bragg grating $\Lambda$ pitch in case of distributed feedback lasers). Generally, $\kappa_\omega > 0$, therefore, the temperature rise in the active layer ($\Delta T > 0$) leads to a negative frequency shift, $\Delta \omega_T < 0$, i.e., it has the opposite sign compared to $\Delta \omega_c$. Thereby, $\Delta \omega_c$ and $\Delta \omega_T$ compete, such that the resulting shift will depend on the values of laser parameters (including parameters defining thermal properties of the laser diode) as well as on the value of $\Delta I$.

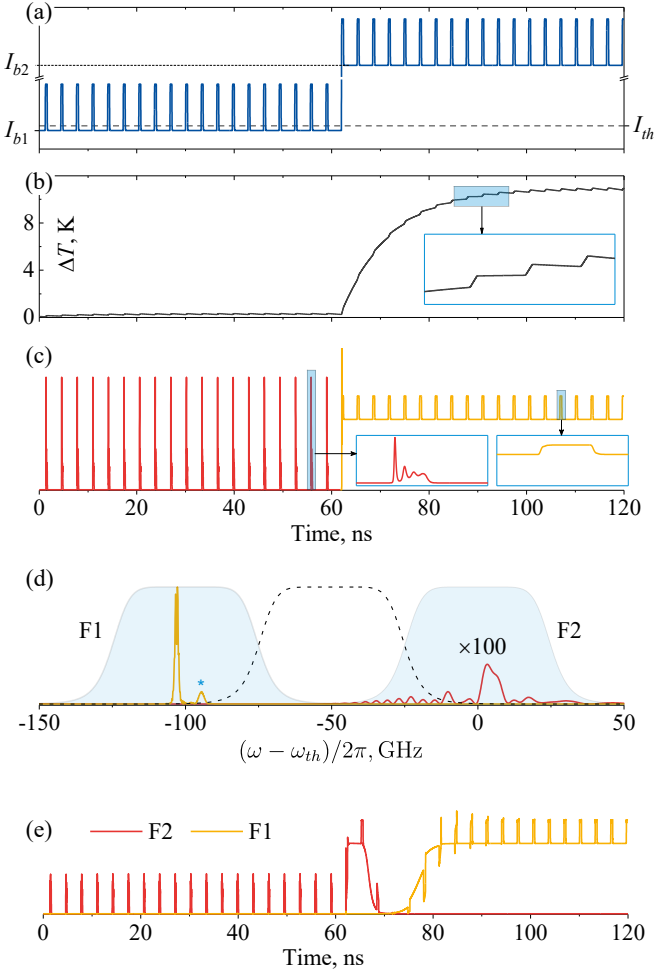

Fig. 4. Simulations of pulses, spectra and temperature dynamics of a laser when switching the pump current. (a) Pump current pulses. Bias currents before and after switching were $I_{b1} = 8$ mA and $I_{b2} = 70$ mA; peak-to-peak value of the modulation current was $I_p = 20$ mA; pulse repetition rate was 312.5 MHz; the pulse width was 0.4 ns. (b) Evolution of the temperature difference $\Delta T$ with time (it was assumed that $\Delta T = 0$ at $I = I_{b1}$). (c) Laser output. (d) Optical spectra plotted with three neighboring transmission windows corresponding to 50 GHz WDM filters. (e) Laser output after splitting into two channels with different 50 GHz bandpass optical filters: the orange line corresponds to the output from filter F1, and the red line corresponds to the output from filter F2. The following parameters were used: $\tau_h = 10$ ns, $r_h = 140$ K/W, $R_L = 8\,\Omega$, $N_{th} = 6.25 \times 10^7$, $N_{tr} = 5.6 \times 10^7$, $\chi_Q = 10^{-6}$, $\tau_{ph} = 2$ ps, $\tau_e = 1$ ns, $\alpha = 5$, $C_{sp} = 10^{-5}$, $\Gamma = 0.2$, $\eta = 0.3$, $\kappa_\omega/2\pi = 12$ GHz.

We define the total shift $\Delta \omega_{tot} = \Delta \omega_c + \Delta \omega_T$, where $\Delta \omega_c$ is estimated from (5), and $\Delta \omega_T = -\kappa_\omega \Delta T$ with the temperature difference $\Delta T$ arises from the heat power $P_a$ dissipated in the active layer as well as from the Joule heat $P_J$ and can be estimated from the following relation:

$$\Delta T = r_h(P_a + P_J) = r_h[V_g(I_2 - I_1)(1 - \eta) + R_L(I_2^2 - I_1^2)], \quad (7)$$

where $r_h$ is an efective thermal resistance of a semiconductor substrate between the active layer and the heat sink, $V_g$ is a voltage across the active layer, and $R_L$ is an effective ohmic resistance of the laser diode. In Fig. 3(a), we show theoretical dependences of the total frequency shift $\Delta \omega_{tot}$ on the pump current $I$ at different values of $r_h$; Fig. 3(b) demonstrates corresponding dependences of $\Delta T$ on $I$ (we put in (5) and (7) $I_1 = I_{th}$ and $I_2 = I$). At small $r_h$ the frequency shift is positive for $I > I_{th}$ (at least in the selected range of pump current variation), indicating that the main contribution to the frequency shift arises from the change of the refractive index. At larger $r_h$, the dissipated power cannot be effectively transferred from the active layer to the heat sink; therefore, the temperature rises much higher than at small $r_h$ making the contribution from heating predominant.

### B. Dynamics of the frequency shift

Theoretical dependences presented above show that for sufficiently large values of $\Delta I$ (and/or with sufficiently high thermal resistance $r_h$), one can switch the lasing frequency between different dense WDM (DWDM) channels by exploiting thermal effects. To be practically applicable for synchronization in QKD, the switching must occur within tens of microseconds and even faster (it is desirable to make the delay $d$ in Fig. 1(b) as short as possible.) Unfortunately, thermal effects can be rather slow, such that the frequency switching may also occur slowly. Nevertheless, in some laser diodes, the temperature may settle down very fast (within tens of nanoseconds) [41].

Several models have been proposed to estimate the transient temperature rise in the active layer induced by the change of the pump current [42-44]. In [40], we used the model of Engeler and Garfinkel [42] to relate material constants of the laser diode to parameters appearing in the phenomenological equation:

$$\frac{d\Delta T}{dt} = -\frac{\Delta T}{\tau_h} + \frac{r_h}{\tau_h}(P_a + P_J), \quad (8)$$

where $\tau_h$ is a thermal rise time. It should be noted that the spread of $\tau_h$ values is very large since it is highly dependent on the structure and materials of the laser diode. In conventional telecom lasers, thermal rise time was measured to be tens of nanoseconds [41], whereas for high-power laser diodes $\tau_h$ of the order of tens and even hundreds of microseconds was reported [45]. The smallness of $\tau_h$ in telecom lasers allows us to assume that switching the laser from the synchronization mode to the regime of quantum state preparation for QKD could be carried out sufficiently fast. Looking ahead, we note that in the telecom lasers that we used in our experiments, the duration of transients was more than hundred of microseconds (not nanoseconds). Nevertheless, this still could be quite fast for some QKD systems.

We performed simulations of laser dynamics when switching the pump current from the value just below threshold to the value much higher than $I_{th}$. For simulations, we used the system (1), where the rate equation for the phase was modified according to (6), and the temperature evolution of the active layer was included according to (8). Simulated pulses, spectra, and temperature dynamics of a laser are shown in Fig. 4. The pump current is shown in Fig. 4(a). The corresponding laser output is shown in Fig. 4(c). Laser pulse shapes before and after the switching are shown in the insets. One can see that before switching, the pulse is accompanied by significant relaxation oscillations, since the laser is assumed to operate in the gain switching mode. After increasing the bias current to $I_{b2}$, the laser operates in the quasi-continuous mode, where relaxation oscillations are almost absent. Note that phase randomization, which is required for QKD and which is performed automatically in a gain-switched laser, is not needed for synchronization, such that we can safely employ the quasi-continuous mode here.

The evolution of the temperature difference $\Delta T$ with time is shown in Fig. 4(b). The inset in Fig. 4(b) shows temperature variation over a short time interval, where step-like jumps of $\Delta T$ are caused by modulation current pulses. Far enough away from the switching point of the bias current, these small variations of $\Delta T$ persist; however, an average temperature does not change.

The spectra shown in Fig. 4(d) are the Fourier transform of a part of simulated data corresponding to single pulses framed in Fig. 4(c). At low bias current, laser pulses are chirped and exhibit broad spectrum (scaled up in Fig. 4(d) for clarity). After increasing the bias current, the spectrum becomes much narrower and is shifted by more than 100 GHz to the left in frequency. Note that it consists of two components: one of them (denoted by an asterisk in Fig. 4(d)) corresponds to the laser pulse itself, whereas the other is related to the "background" radiation. These components are shifted relative to each other by a few gigahertz due to the chirp effect. The frequency difference between the two components can be estimated by substituting $\Delta I = I_p$ in (5).

Presented simulations show that the lasing frequency can be tuned by tens of GHz making it possible to switch between different WDM channels (for illustration, we plotted in Fig. 4(d) transmission windows corresponding to 50 GHz DWDM filters). In Fig. 4(e), we show simulations of the same laser output as in Fig. 4(c), but now after splitting the signal into two channels with different 50 GHz bandpass optical filters (F1 and F2 in Fig. 4(d)). In contrast to Fig. 4(c), where the switching of the bias current occurs almost without transients, the filtered signal in Fig. 4(e) exhibits significant transients, whose duration is determined by the thermal rise time.

## IV. EXPERIMENT

### A. Measurements of $\kappa_\omega$ and $r_h$

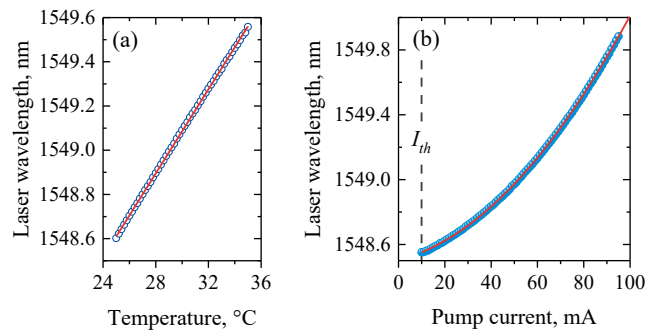

Fig. 5. Experimantal dependences of the laser wavelength on the laser temperature (a) and on the pump curent (b). Solid lines fit to experimental data.

Before conducting an experiment on TWDM via pump current variation in a pulsed laser, we measured the dependences of the lasing wavelength on laser temperature and on the pump current in the CW mode (see Fig. 5). For these measurements, we used a NOLATECH 1550 nm distributed feedback (DFB) laser in a butterfly package, which was driven by the Thorlabs CLD1015 laser driver equipped with the temperature controller. For each temperature value, the wavelength was determined from the peak position of the spectral line, which was measured with Yokogawa AQ6370D optical spectrum analyzer. The temperature dependence of the laser wavelength is shown in Fig. 5(a). It provides the value of the temperature coefficient of the wavelength $\kappa_\lambda$, which was estimated from the slope of the straight line that fits experimental data. We found $\kappa_\lambda = 0.095$ nm/K; to recalculate it into $\kappa_\omega$ we used the relation $\kappa_\omega = 2\pi c \kappa_\lambda / \lambda_{th}^2$, where $c$ is the speed of light in vacuum, and the wavelength $\lambda_{th} = 1548.55$ nm was measured at threshold current $I_{th}$ at temperature 20°C (the value of $I_{th}$ was estimated from the LI-curve and was found to be 10 mA.) We obtained $\kappa_\omega / 2\pi = 11.9$ GHz. (Note that both $\kappa_\lambda$ and $\kappa_\omega$ should be positive since the minus sign in front of $\kappa_\omega$ has been taken into account in (6).)

To fit the dependence of the lasing wavelength on the pump current, we used the relation $\lambda(I) = 2\pi c / \omega(I)$, where the steady-state lasing frequency is defined by

$$\omega(I) = \omega_{th} + \frac{\alpha A}{2\tau_{ph}} - \kappa_\omega \Delta T \quad (9)$$

with

$$A = B - 1 - q\chi_Q,$$
$$B = \left[1 + \chi_Q^2 q^2 + 2\frac{\Gamma\tau_{ph}}{e}(I - I_{tr})\chi_Q\right]^{1/2}, \quad (10)$$
$$q = \frac{\Gamma\tau_{ph}}{e}(I_{th} - I_{tr}).$$

Relations (9)–(10) have been obtained from the system (1), where we neglected the spontaneous emission term ($C_{sp} = 0$) and put $\dot{Q} = \dot{N} = 0$ and $\dot{\varphi} = \omega - \omega_{th}$ with $\omega_{th} = 2\pi c/\lambda_{th}$. The temperature difference in (9) was determined by (7). We fixed most of the parameters for $\omega(I)$ with only two of them left to fit: thermal resistance $r_h$ and effective ohmic resistance $R_L$. The parameters $\tau_{ph}$, $\tau_e$, $\alpha$, $\chi_Q$, $\eta$, and $\Gamma$ were the same as in simulations in Fig. 4. The current at transparency was set to $I_{tr} = 0.9 I_{th}$, and the voltage across the active layer was assumed to be $V_g = 0.8$ V. Due to this quite arbitrary choice of parameters, the obtained values of $r_h$ and $R_L$ should be treated as a relatively rough estimate rather than the direct measurement. However, if all the predefined parameters were found experimentally, the fitting with (9) could be considered to be a sufficiently rigorous approach to determine $r_h$ and $R_L$. The fitting curve shown in Fig. 5(b) corresponds to the values $r_h = 147$ K/W and $R_L = 7.9\,\Omega$, which are in a good agreement with theoretical predictions shown in Fig. 3 and also with values of thermal resistance measured experimentally in conventional telecom lasers by the so-called TRAIT method [46-48].

### B. Proof-of-principle demonstration

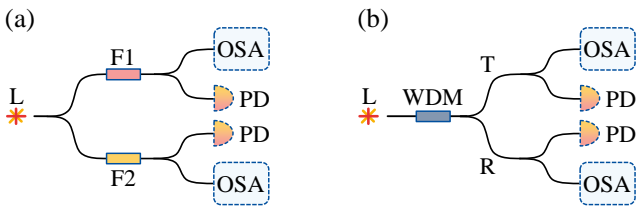

Fig. 6. Simplified optical schemes to perform proof-of-principle demonstration of time- and wavelength-division multiplexing via pump current variation in a pulsed laser. R and T stand for reflection and transmission arms of the WDM filter, respectively.

To demonstrate the TWDM discussed in section III, we used the same NOLATECH 1550 nm DFB laser, for which we performed measurements of $\kappa_\omega$ and $r_h$ in the previous section. We also used the Shengshi 1550 nm DFB laser, with similar specifications, but which allows a higher modulation frequency. We used the first laser at a modulation frequency of 312.5 MHz, whereas the second one at 2.5 GHz. For modulation, we used a home-build pulsed laser driver based on the ONET1151L 11.3-Gbps low-power laser-diode driver chip. Thermal stabilization was performed by commercially available single-chip temperature controller (thermoelectric cooler is included by default into a standard butterfly package). The waveforms modulated at 312.5 MHz and 2.5 GHz were generated by a phase-locked loops multiplying the input frequency from the 10 MHz reference oscillator.

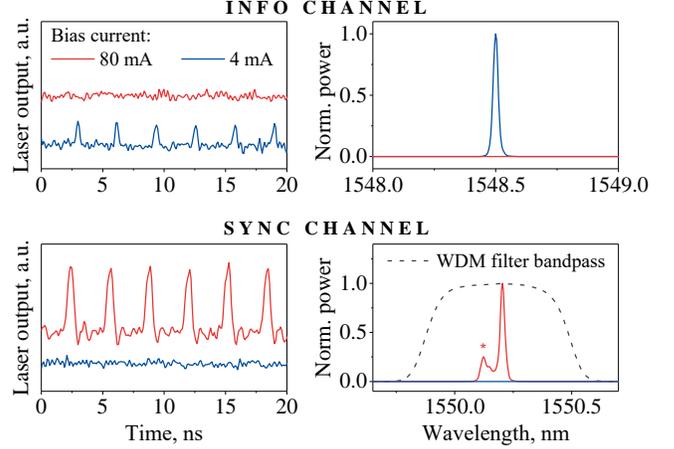

Fig. 7. Laser pulses (on the left) and corresponding spectra (on the right) in information and synchronization channels measured at pulse repetition rate of 312.5 MHz with the optical scheme presented in Fig. 6(a).

First, we measured laser pulses and corresponding spectra in information and synchronization channels at a pulse repetition rate of 312.5 MHz (Fig. 7) with a NOLATECH laser using the optical scheme presented in Fig. 6(a). The laser output was split into two channels, which we call the "Info channel" (for definiteness, let it be a channel with the filter F1) and the "Sync channel" (a channel with the filter F2). The Santec OTF-980 optical tunable filter (the bandpass was put to 6.25 GHz) was used as F1 and a standard 100 GHz DWDM filter (C34) as F2. The laser was modulated by rectangular current pulses of (nominal) width 800 ps; the peak-to-peak value of the modulation current was estimated to be approximately 80 mA. Laser pulses entered the information channel (i.e., they passed through the filter F1) when the bias current $I_b$ was below threshold ($I_{b1} = 4$ mA), whereas the value $I_{b2} = 80$ mA was used to emulate the synchronization mode. One can see from Fig. 7 that such a difference between the bias currents ($I_{b2} - I_{b1} = 76$ mA) yields the spectral shift of more than 1.6 nm (200 GHz), which allows separating information and synchronization signals with negligible crosstalk. Note that in agreement with simulations in Fig. 4(d), the laser spectrum at $I_{b2} = 80$ mA consists of two components (the one that corresponds to the laser pulse itself is denoted by an asterisk). For clarity, the WDM filter passband is shown together with the laser spectrum in the sync channel by the dashed line. Note that the passband of the filter F1 coincides with the spectrum shown in Fig. 7 (for "info channel") since we cut out a narrow spectral component (6.25 GHz) from a wide spectrum of chirped pulses.

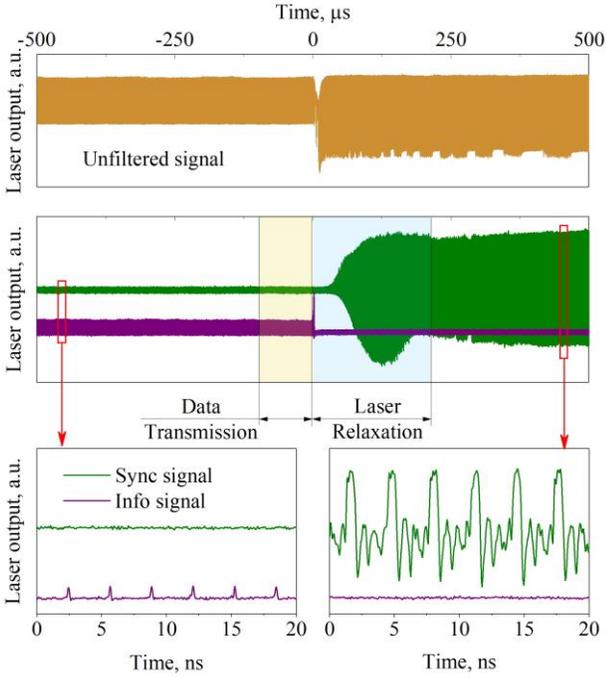

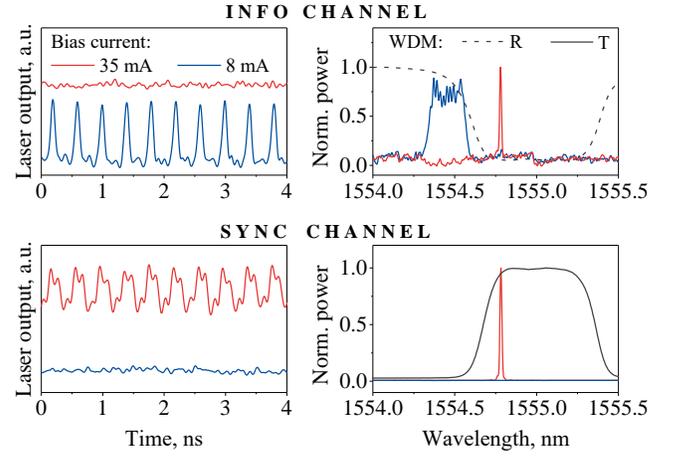

Fig. 8. Laser pulse train before and after changing the bias current from $I_{b1} = 4$ mA to $I_{b2} = 80$ mA recorded at a pulse repetition rate of 312.5 MHz without spectral filtering (at the top). Corresponding pulse trains (synchronization and information signals) recorded with the optical scheme presented in Fig. 6(a) (in the middle).

Figure 8 demonstrates transients occurring when switching the laser from the information to the synchronization mode. At the top of figure, we show the signal acquired without filters F1 and F2. The transients observed in this signal correspond to relaxation processes occurring in the laser driver, which, as can be seen, are relatively short. In the middle of Fig. 8, we plotted the signal recorded with filters F1 and F2 (the same as in Fig. 7) during 1 ms. The light-yellow rectangle marks the period during which data is transferred between the processing unit (field-programmable gate array in our case) and the laser driver. The light-blue rectangle marks the period approximately corresponding to the duration of temperature relaxation processes in a laser. The former was estimated to be around 100 µs, whereas the latter was revealed to be at least 200 µs. Laser pulses in the nanosecond range are shown at the bottom of Fig. 8.

To carry out the experiment at pulse repetition rate of 2.5 GHz, we used the Shengshi 1550 nm DFB laser and the scheme shown in Fig. 6(b). Laser pulses and corresponding spectra at $I_{b1} = 8$ mA (just below threshold) and $I_{b2} = 35$ mA are shown in Fig. 9. We intentionally used a smaller difference $I_{b2} - I_{b1}$ in this experiment to clearly demonstrate that the scheme in Fig. 2(b) may be insecure in the context of QKD (we will discuss this below). The laser spectra at chosen bias currents are only 0.32 nm apart (approximately 40 GHz), which does not allow separating the synchronization and information signals into different WDM channels. However, the spectrum of the synchronization signal can be placed at the edge of the WDM filter's passband, allowing for efficient separation of the signals, as shown in Fig. 9.

Fig. 9. Laser pulses (on the left) and corresponding spectra (on the right) in information and synchronization channels measured at pulse repetition rate of 2.5 GHz with the optical scheme presented in Fig. 6(b). R and T stand for reflection and transmission arms of the WDM filter, respectively.

Figure 10 demonstrates laser pulse trains before and after changing the bias current from $I_{b1} = 16$ mA and $I_{b2} = 35$ mA recorded at a pulse repetition rate of 2.5 GHz with the optical scheme presented in Fig. 6(b). In this experiment, laser relaxation occurred faster than in the experiment at 312.5 MHz. We believe that such a difference in the duration of the relaxation process was due to lower $I_{b2} - I_{b1}$ value (19 mA vs 74 mA) rather than to the difference in laser diodes. Indeed, we observed quite different durations of temperature transients in both lasers at different $I_{b2} - I_{b1}$ values (not presented in this work).

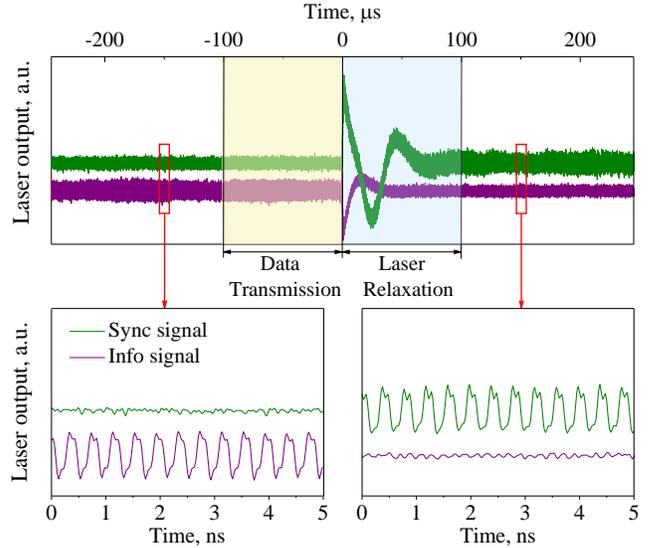

Fig. 10. Laser pulse trains before and after changing the bias current from $I_{b1} = 16$ mA to $I_{b2} = 35$ mA recorded at a pulse repetition rate of 2.5 GHz with the optical scheme presented in Fig. 6(b).

## V. DISCUSSION

The first thing we would like to note is the discrepancy between the relaxation times in simulations, where we put a thermal rise time to be 10 ns (see Fig. 4(e)), and in the experiment, where we observed temperature transients lasting 100–200 μs. In fact, the temperature rise time can vary over a very wide range in real laser diodes, so the observed discrepancy is quite expected. What was unexpected for us, however, was the dependence of the duration of temperature transients on the difference in bias currents $I_{b2} - I_{b1}$. Such dependence indicates that the temperature dynamics in a laser (at least at relatively large values of $I_{b2} - I_{b1}$) seems to be different from a simple exponent $\exp(-t/\tau_h)$. In other words, a simple model of temperature dynamics in a semiconductor laser given by (8) and discussed, e.g., in [40] should be used with caution.

Another important note, now related to QKD, is that although optical schemes shown in Fig. 2 are functionally equivalent, they still have a significant difference in terms of security. In order to demonstrate this difference, we performed an experiment at a pulse repetition rate of 2.5 GHz using the optical scheme in Fig. 6(b) (which is the proof-of-principle analog of the scheme in Fig. 2(b)). The main disadvantage of the scheme in Fig. 2(b) is that the sync channel bypasses the attenuator (which is assumed to be a part of the encoder in Fig. 2), so any spectral components leaked into this channel (in particular, obtained because of re-reflections in active optical components) will be also sent to Bob, duplicating the main signal. In fact, one can see from Fig. 9 that the spectrum of the info signal is quite broad, such that it partially enters the passband. (This is clearly seen even in the linear plot; in the logarithmic scale, the situation looks even worse.) Obviously, this can lead to information leakage, so this scheme should be used carefully. For the scheme in Fig. 2(a), this problem is leveled by the fact that the filter F2 can always be made sufficiently narrow-band and placed far enough away from the information signal on the frequency scale, so that almost no spectral components of the prepared quantum states will pass through it.

We also note that there is no fundamental difference between the experiments at different laser pulse repetition rates and one may employ either in Fig. 2(a) or in Fig. 2(b) at any frequency. However, it should be remembered that the risk of information leakage through the reflection channel of the WDM filter remains even at large values of $I_{b2} - I_{b1}$, i.e., at a significant spectral shift between the sync and info signals.

Finally, we would like to answer the question that may seem strange in the light of this work: is it mandatory to use WDM in the synchronization scheme with a single laser? The answer may also seem strange: in principle, WDM is not necessarily needed! Indeed, it is sufficient to install a beam splitter (or an optical switch) on Bob's side in front of the single photon detector to divert part of the signal to the sync detector. However, in this case (when using, e.g., a 50:50 beam splitter), we increase the effective channel loss, which may lead to a significant reduction in key generation rate. One can try to level this problem by using an asymmetric beam splitter diverting most of the signal to the single photon detector; however, at sufficiently large distances, the signal coming to the sync detector may be too weak and cannot be thus used for synchronization. In addition, in the optical scheme without WDM (it will be almost identical to the schemes in Fig. 2), an amplitude modulator should be used instead of the optical attenuator since one should switch the channel from info to sync mode quite frequently. Obviously, this will lead to additional complication and cost of the system. Thus, a single laser synchronization scheme using WDM seems to be more attractive.

## VI. Conclusions

We studied in detail the change in the lasing wavelength of a semiconductor laser with a change in the pump current and clearly showed that when thermal resistance of a laser diode is high, the lasing wavelength is red-shifted due to the change of the refractive index caused by a thermal effect and this shift may significantly exceed (in absolute value) the blue shift related to a refractive index change induced by the carrier injection. We have shown that by measuring the dependence of the laser wavelength on the pump current and using a simple thermal model of a laser diode, one can estimate its thermal resistance. Using the obtained results, we proposed a method of synchronization for QKD systems using wavelength- and time-division multiplexing via pump current variation of a pulsed laser. To demonstrate the proposed method, we performed numerical simulations as well as proof-of-principle experiments at 312.5 MHz and 2.5 GHz of laser pulse repetition rate. Temperature transients were revealed to be fast enough, which allows using our method in real QKD systems. We believe that the theoretical and experimental results obtained here could be useful for the developers of these systems.

## Acknowledgment

We are grateful to Akky Feimov and Vasily Yashchuk for valuable comments.